\documentclass[11pt]{article}
\usepackage{moriond,epsfig}

\def\beq{\begin{equation}}
\def\eeq{\end{equation}}
\def\beqa{\begin{eqnarray}}
\def\eeqa{\end{eqnarray}}

\def\pref#1{(\ref{#1})}

\begin{document}
\vspace*{4cm}
\title{SUPERSYMMETRIC LARGE EXTRA DIMENSIONS}

\author{ C.P. BURGESS }

\address{Physics Department, McGill University, 3600 University Street,\\
Montr\'eal, Qu\'ebec, Canada, H3A 2T8}

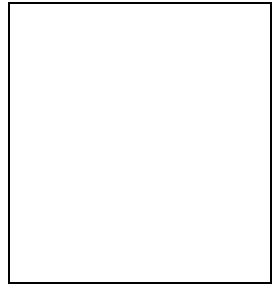
\maketitle\abstracts{ This article summarizes the proposal to
address the cosmological constant problem within the framework of
supersymmetric large extra dimensions (SLED). The proposed
mechanism is described, emphasizing the relaxation mechanism which
ensures that low-energy particles like the electron do not
contribute too large a vacuum energy. This is followed by a
preliminary discussion of outstanding issues and observational
consequences.}

\section{Introduction}

Current measurements indicate that the universal expansion is
currently accelerating in a way which is consistent with the
existence of a small but nonzero cosmological vacuum energy
density which is of order $\rho = v^4$, with $v \sim 3 \times
10^{-3}$ eV in units for which $\hbar = c = 1$. This discovery is
a challenge because all known theories of microscopic physics
appear to predict that a particle of mass $m$ contributes to
$\rho$ an amount which is of order $\delta \rho \sim m^4$, and
almost all known elementary particles have masses much larger than
$10^{-3}$ GeV. This is particularly disturbing because it
indicates a problem with our description of {\it low energy}
physics --- such as the physics of the electron (for which $m_e =
5 \times 10^5$ eV) --- which we normally think we understand.

Supersymmetry is one of the few symmetries which may be helpful.
This is because sufficiently many supersymmetries can explain
forbid a cosmological constant in the theory at very short
distances. Furthermore, even 4D $N=1$ supersymmetry can also
partially explain why the process of integrating out particles
does not produce too large a vacuum energy, since supersymmetry
enforces a cancellation between the contributions of bosons and
fermions. Unfortunately, this cancellation is only partial if
supersymmetry is spontaneously broken, leaving a residual value
which must vanish either as $m_{sb}^2 \, M^2$ or $m_{sb}^4$, where
$M$ is the largest mass scale in the problem and $m_{sb}$ is the
supersymmetry-breaking scale. Unfortunately, the absence of
observed superpartners for particles like the electron already
implies that $m_{sb}$ must be at least as large as hundreds of
GeV.

\subsection{Supersymmetric Large Extra Dimensions (SLED)}

In the SLED proposal \cite{branesphere} it is supposed that
progress can be made in understanding the observed size of the
vacuum energy within the framework of supersymmetric large extra
dimensions. In this picture the world has two unwarped extra
dimensions whose radii satisfy $1/r \sim v \sim 10^{-2}$ eV --- in
which case $r \sim 1/v \sim 0.01$ mm --- such as has been proposed
in the Large Extra Dimensional (LED) scenario.\cite{add}
Dimensions this large can have escaped observational detection
provided all of the known particles (except the graviton) are
trapped on a 4-dimensional surface (3-brane) which is localized
within the two extra dimensions. Dimensions this large have
previously been motivated to understand the hierarchy problem,
since they predict that the effective 4D Planck scale, $M_p$, is
related to the underlying scale of 6D gravity, $M$, by the
relation $M_p = M^2 \, r$. If $1/r \sim 0.01$ eV then the correct
size, $M_p \sim 10^{19}$ GeV, is obtained for $M$ near the weak
scale: $M \sim 10$ TeV.\footnote{We use here the Jordan frame, for
which $M_p$ is $r$ dependent but the electroweak scale, $M$, is
not. It is in this frame that the Kaluza-Klein mass scale is
$m_{\rm KK} \sim 1/r$. The constraint $r < 0.01$ mm comes from the
requirement of not having excessive energy loss from astrophysical
objects like supernovae.\cite{LEDastrobounds,susyaddbounds}}

The difference between the SLED and LED proposals is that the
theory is also required to be supersymmetric, as might naturally
be expected if it were to arise as a consequence of an underlying
theory such as string theory. In this case the absence of
supersymmetry in collider experiments can be understood if
supersymmetry is strongly broken on our brane, perhaps at the
scale $M \sim 10$ TeV. The main requirement of the SLED proposal
is that the supersymmetry breaking scale in the 6D bulk is the
natural size which would be inherited due to its gravitational
strength couplings to the branes, that is $m_{sb} \sim M^2/M_p
\sim 1/r \sim v \sim 0.01$ eV. In this case the observed vacuum
energy density is close to the supersymmetry-breaking scale in the
bulk.

\section{The Cosmological Constant Problem Seen from 6 Dimensions}

This section describes the contributions to the observed vacuum
energy which can be expected within the SLED framework. In order
to do so we first imagine integrating out all brane modes; and
then integrating out all of the bulk physics to obtain the
effective vacuum energy on scales much larger than the
extra-dimensional size, $r$.

\subsection{Dynamical Relaxation and the Cancellation of Brane
Tensions}

One of the biggest problems for understanding the size of the
vacuum energy has always been how to understand why
well-understood particles like the electron do not contribute too
large an amount. Since supersymmetry is badly broken on the brane
within SLED, integrating out the known particles (like the
electron) indeed does contribute a large vacuum energy density,
which in total is naturally of order $M^4$. But this energy
density is {\it not} a cosmological constant. Rather, it is a
contribution to the tension of the brane --- $\delta T \sim M^4$
--- and so is an energy source which is localized within the
extra dimensions at the position of the brane. We must ask how
this energy source curves the extra dimensions, and then see what
the implications are for the effective 4D cosmological constant
which would be observed on distance scales larger than $r \sim
0.1$ mm.

It happens that the curvature of the extra dimensions due to this
localized energy source can be computed using the classical
equations of motion in the bulk, with the result that the geometry
acquires a conical singularity at the position of the branes. This
singularity contributes a delta-function contribution to the
two-dimensional curvature given by
\beq \label{Ricci2}
    R_2 = - {2\over e_2} \, \sum_i \left( \frac{T_i}{M^4} \right) \,
    \delta^2(y-y_i) + \dots \, ,
\eeq
where $e_2$ is the volume element for the internal two dimensions,
$y_i$ denotes the position of the `$i$'th brane and the ellipses
denote contributions to $R_2$ which are smooth at the position of
the brane.

The leading contribution to the effective 4D cosmological constant
on long distance scales is obtained by integrating out the bulk
gravitational degrees of freedom at the classical level. In this
result an interesting cancellation occurs between the brane
tensions and the classical bulk curvature given by
eq.~\pref{Ricci2}. The effective 4D cosmological constant obtained
at this order is
\beq \label{rhocl}
    \rho_{\rm cl} =  \sum_i T_i + \int_M d^2y \; e_2 \,
    \left[\frac{M^4}{2} \, R_2 + \dots \right] \nonumber \\
    = 0 \, ,
\eeq
where the sum on `$i$' is over the various branes in the two extra
dimensions and `$\dots$' denotes all of the other terms besides
the Einstein-Hilbert term in the supersymmetric bulk action.
Interestingly the sum over brane tensions, $T_i$, precisely
cancels the contribution of the singular part of the curvature,
eq.~\pref{Ricci2}, to which they give rise,\cite{clp} and for
supersymmetric theories the same kind of cancellation also occurs
amongst the remaining terms in $\rho_{\rm cl}$ once these are
evaluated using the smooth parts of the geometry and the other
bulk fields obtained using the classical field
equations.\cite{branesphere} This cancellation has its roots in a
classical scale invariance of the 6D supergravity
action.\cite{susyads}

The above arguments have recently been verified in a very explicit
way, through the construction of a very wide class of solutions to
the classical field equations of 6D supergravity.\cite{ggp} The
authors of this reference construct the most general 2-brane
solutions which are ($i$) maximally symmetric in the 4 large
dimensions, ($ii$) axially symmetric in the 2 extra dimensions,
and ($iii$) for which the 6D dilaton --- a 6D superpartner to the
metric --- and warp factor are nonsingular at the position of the
two branes.\footnote{This last condition is equivalent to choosing
the brane-dilaton coupling to preserve the classical scale
invariance of the bulk supergravity action.\cite{susyads}} The
intrinsic geometry of the large 4 dimensions is flat for every
single one of these solutions, as the above arguments would
require. In this way we see very robustly how the classical bulk
response can systematically cancel the contributions of any brane
tensions to the effective 4D vacuum energy.

\subsection{Bulk Loops}

The above is a classical argument within the bulk, and so a
central question asks for the size of quantum corrections. It is
now argued that these can be acceptably small provided that the
supersymmetry breaking scale in the bulk is $m_{sb} \sim 1/r \sim
v$, as argued earlier. As we have seen, a contribution of the form
$\delta \rho \sim m_{sb}^4$ gives precisely the observed value
when $m_{sb}$ is this large.

Although $m_{sb}^4$ can arise for the vacuum energy in a
supersymmetric theory, it is not automatic since supersymmetry
just requires the result to vanish as $m_{sb} \to 0$. In
particular, it does not preclude the appearance of contributions
of order $m_{sb}^2 M^2$. Such terms also have a natural
interpretation in 6 dimensions as arising as local
curvature-squared contributions to the effective 6D action as
high-energy modes having wavelength $\lambda \sim 1/M$ are
integrated out.\cite{branesphere,update} Although such
contributions can arise for some 6D supergravities, they need not
for all supergravities.

Detailed calculations to determine the size of the quantum
contributions are underway, and at present suggest that several
conclusions may be drawn. In general there appear to be two
reasons why such curvature-squared terms do not contribute for
some 6D supergravities. Their presence can be forbidden by the
existence of additional supersymmetries at the TeV scale $M$, such
as what would arise if the 6D theory were to arise as the
low-energy limit of a higher-dimensional supergravity at these
scales. Alternatively, even if they arise these local
curvature-squared terms may vanish once evaluated at the classical
solution, such as would occur for flat toroidal solutions. Since
the 6D classical field equations can tie the value, $\phi$, of the
6D dilaton to the size of the extra dimensions by $e^\phi \sim
1/r^2$, it can also happen that it is sufficient to have the
dangerous $m_{sb}^2 M^2$ terms vanish only for a few orders in the
loop expansion (such as one loop) in order to obtain a
satisfactorily small vacuum energy.\cite{update}

Indirect evidence that ultraviolet extensions of 6D supergravity
exist for which the loop corrections are of order $\rho \sim
m_{sb}^4$ comes from direct one-loop string-based calculations of
the vacuum energy. These calculations have been performed for
toroidal compactifications with supersymmetry broken by boundary
conditions and/or the presence of branes,\cite{ablm} and give a
result which is of order $\rho \sim m_{sb}^4 \sim 1/r^4$.

\subsection{Topological Constraints}

The ability to identify broad classes of solutions in 6D
supergravity\cite{ggp,susyads} allows one to see whether or not
the classical cancellation of the vacuum energy has somehow been
ensured through a hidden fine-tuning of some of the parameters of
the theory. This kind of hidden fine-tuning has been been raised
as an objection \cite{5dtune} to the 5D models of
ref.~\cite{5dst}. Examination of the general solutions shows that
for all of them the brane tensions (and other parameters) indeed
satisfy two constraints. These constraints simply express the
topological statement that the Euler number of the internal
geometry is the same as for a 2-sphere, and that the background
Maxwell fields in the solutions have the Chern class appropriate
to a magnetic monopole.\cite{ggp,susyads} These constraints should
be regarded as being topological {\it integrability conditions}
which must be satisfied in order for a solution to the classical
field equations to exist, but for {\it all} choices for which
solutions exist the intrinsic 4D geometry is flat.

As far as naturalness is concerned, the central point is whether
these constraints remain satisfied as successive scales are
integrated out from the high scale, $M \sim 10$ TeV, to the low
scale $1/r \sim 10^{-2}$ eV. The topological constraints are
expected to be stable against integrating out the scales between
$M$ and $1/r$ precisely because they are topological. Being
topological, they state that a particular combination of brane and
bulk quantities is an integer and so cannot renormalize as
short-distance modes are integrated out.\cite{susyads,update}

The topological constraints do contain a danger, however, since
they show is that for generic tensions the internal 2D geometry
{\it warps}. This is dangerous from the present perspective
because a vacuum energy of order $m_{sb}^4$ is only small enough
if $m_{sb} \sim 1/r \sim 0.01$ eV. But the value of $r$ is also
fixed by the hierarchy problem, and typically only takes the
numerical value 0.01 eV in the unwarped case\cite{susyads}. This
is because warping in general also contributes to the hierarchy
--- {\it \`a la} Randall and Sundrum\cite{RS} --- and so can allow the
Kaluza Klein masses to be larger than in the unwarped case.

\section{Observational Consequences}

Although it is motivated by the cosmological constant problem, the
SLED proposal has many rich experimental implications: for
cosmology, for tests of gravity and for high-energy accelerators.
Many of these implications are interesting in their own right,
since they are very different from the usual paradigm of
weak-scale supersymmetry breaking which leads to the
Supersymmetric Standard Model. This section closes with a summary
of some of these potential implications.\cite{msled}

\begin{itemize}

\item {\bf Late-Time Cosmology} As might be expected for a
relaxation mechanism, given that the extra dimensions are
presently $r \sim 0.01$ mm in size the 6D SLED proposal implies
the existence of a very light scalar field whose mass at present
is of order the present-day Hubble scale: $m_{\varphi} \sim H_0
\sim (M_p r^2)^{-1} \sim 10^{-33}$ eV. It is remarkable that such
a small scalar mass is stable against radiative corrections
provided that the corrections to the 4D vacuum energy are of order
$1/r^4$, as argued above.\cite{ABRS2,update} As a result the
phenomenology of the Dark Energy naturally takes a `quintessence'
form, with an Albrecht-Skordis potential.\cite{AS,ABRS2} Studies
are underway to determine whether such cosmologies can satisfy
observational constraints, such as the limits on the Dark Energy
equation of state and limits on the evolution of Newton's
constant.

\item {\bf Early-Time Cosmology} LED models are known to be
subject to strong constraints, such as due to the requirement that
there not be too much population of bulk modes during
nucleosynthesis.\cite{add,LEDcosmobounds} SLED models typically
feel these bounds even more strongly\cite{susyaddbounds} due to
the larger number of states which are available in the bulk. In
particular it is not yet clear how weak-scale cosmology proceeds
in these models, including issues such as the origin within SLED
of Dark Matter.\cite{update}

\item {\bf Long-Range Tests of Gravity} The existence of such a
light scalar implies the existence of a very long-range force
which acts in competition with gravity. Strong limits exist on
such forces\cite{LRTests} which provide potentially severe
constraints on 6D SLED models. A potential loophole which these
models may use to thread these bounds relies on the
time-dependence of the scalar couplings to ordinary
matter.\cite{ABRS2}

\item {\bf Astrophysical Constraints} 6D LED models are known to
be subject to strong constraints from supernovae and astrophysics,
which constrain the rate with which such systems can drain their
energy by radiation into the extra
dimensions.\cite{LEDastrobounds} This is the origin of the
estimate that $M \sim 10$ TeV (and hence $r \sim 0.01$
mm).\cite{msled}

\item {\bf Short-Range Tests of Gravity} LED and SLED models
predict deviations to Newton's Law at sub-millimetre distances due
to the opening up there of the extra dimensions.\cite{NLTests}
However for SLED the successful description of the vacuum energy
density fails if $r$ becomes much smaller than 0.01 mm, and so
there is a definite prediction of the scale at which new phenomena
{\it must} become visible. The predicted range for deviations from
Newton's Law is of order $r/(2\pi) \sim 1 \; \mu$m, and shorter
distances cannot be entertained without giving up the explanation
for the vacuum energy.

\item {\bf Accelerator Physics} Since gravitational physics must
be at the TeV scale, there should be observable signals of
extra-dimensional gravity at TeV-scale accelerators like the Large
Hadron Collider just as for the non-supersymmetric LED picture.
Although detailed modelling of the resulting signatures is in its
infancy,\cite{susyaddbounds,susyaddbounds2} the resulting
missing-energy signals would look very different from what would
be expected for supersymmetric extensions of the Standard
Model.\cite{msled}

\end{itemize}

\section*{Acknowledgements}

This work was supported in part by funds from NSERC (Canada), FCAR
(Qu\'ebec) and McGill University, and arose out of collaborations
with Yashar Aghababaie, Jim Cline, Hassan Firouzjahi, Susha
Parameswaran, Fernando Quevedo, Gianmassimo Tasinato and Ivonne
Zavala. I'd like to thank the organizers of the Moriond
Electroweak Workshop for their kind invitation to speak, as well
as for providing such a first-class environment for the meeting.


\begin{thebibliography}{99}

\bibitem{branesphere}
Y. Aghababaie, C.P. Burgess, S. Parameswaran and F. Quevedo,
Nucl.~Phys.~{\bf B} (to appear), (hep-th/0304256).

\bibitem{add}
N. Arkani-Hamed, S. Dimopoulos and G. Dvali, { Phys.\ Lett.} {\bf
B429} (1998) 263  hep-ph/9803315;
%
Phys.\ Rev.\ {\bf D59} (1999) 086004  [hep-ph/9807344];
%
I.~Antoniadis, N.~Arkani-Hamed, S.~Dimopoulos and G.~R.~Dvali,
Phys.\ Lett.\ B {\bf 436} (1998) 257 [hep-ph/9804398].

\bibitem{LEDastrobounds} S. Cullen and M. Perelstein, {\it Phys.
Rev. Lett.} {\bf 83} (1999) 268 [hep-ph/9903422];
%
V. Barger, T. Han, C. Kao, R.-J. Zhang, Phys.\ Lett.\ {\bf B461}
(1999) 34-42, [hep-ph/9905474];
%
C. Hanhart, D.R. Phillips, S. Reddy, M.J. Savage, {\it Nucl.
Phys.} {\bf B595} (2001) 335 [nucl-th/0007016];
%
J.A. Pons, D.R. Phillips and S. Reddy, Phys.\ Lett.\ {\bf B509}
(2001) 1-9, [astro-ph/0102063];
%
S. Hannestad and G. Raffelt, Phys.\ Rev.\ Lett.\ {\bf 87} (2001)
051301, [hep-ph/0103201]; Phys.\ Rev.\ Lett.\ {\bf 88} (2002)
071301, [hep-ph/0110067]; Phys.\ Rev.\ {\bf D67} (2003) 125008,
Erratum-ibid.\ {\bf D69} (2004) 029901, [hep-ph/0304029];
%
S. Hannestad, Phys.\ Rev.\ {\bf D64} (2001) 023515,
[hep-ph/0102290].

\bibitem{susyaddbounds}
D. Atwood, C.P. Burgess, E. Filotas, F. Leblond, D. London and I.
Maksymyk, ``Supersymmetric Large Extra Dimensions are Small and/or
Numerous'', Phys.~Rev. {\bf D63} (2001) 025007, (hep-ph/0007178).

\bibitem{clp}
J.-W. Chen, M.A. Luty and E. Pont\'on, JHEP 0009 (2000) 012,
[hep-th/0003067];
%
S.M. Carroll and M.M. Guica, [hep-th/0302067];
%
I. Navarro, JCAP 0309:004,2003, [hep-th/0302129].

\bibitem{susyads}
Y. Aghababie, C.P. Burgess, J.M. Cline, H. Firouzjahi, F. Quevedo,
G. Tasinato and I. Zavala, JHEP 0309 (2003) 037, (hep-th/0308064).

\bibitem{ggp}
G.W.~Gibbons, R.~G\"uven and C.N.~Pope, [hep-th/0307238].

\bibitem{update}
C.P. ~Burgess, Ann. Phys. (NY) (to appear) [hep-th/0402200].
%

\bibitem{ablm}
See for instance: I.~Antoniadis, K.~Benakli, A.~Laugier and
T.~Maillard, Nucl.\ Phys.\ B {\bf 662} (2003) 40
[arXiv:hep-ph/0211409];
%
M.~Klein, Phys.\ Rev.\ D {\bf 67} (2003) 045021
[arXiv:hep-th/0209206];
%
C.~Angelantonj and I.~Antoniadis, [hep-th/0307254].

\bibitem{5dtune}
S.~Forste, Z.~Lalak, S.~Lavignac and H.~P.~Nilles, Phys.\ Lett.\ B
{\bf 481} (2000) 360, hep-th/0002164; JHEP {\bf 0009} (2000) 034,
[hep-th/0006139];
%
C.~Csaki, J.~Erlich, C.~Grojean and T.J.~Hollowood, Nucl.\ Phys.\
{\bf B584} (2000) 359-386, [hep-th/0004133];
%
C.~Csaki, J.~Erlich and C.~Grojean, Nucl.\ Phys.\ {\bf B604}
(2001) 312-342, [hep-th/0012143];
%
J.M. Cline and H. Firouzjahi, Phys.\ Rev.\ {\bf D65} (2002)
043501, [hep-th/0107198].


\bibitem{5dst}
N.~Arkani-Hamed, S.~Dimopoulos, N.~Kaloper and R.~Sundrum,
Phys.\ Lett.\ B {\bf 480} (2000) 193, [hep-th/0001197];
%
S.~Kachru, M.~B.~Schulz and E.~Silverstein,
Phys.\ Rev.\ D {\bf 62} (2000) 045021, [hep-th/0001206].

\bibitem{RS}
L. Randall and R. Sundrum, Phys.\ Rev.\ Lett.\ {\bf 83} (1999)
3370-3373, [hep-ph/9905221]; Phys.\ Rev.\ Lett.\ {\bf 83} (1999)
4690-4693, [hep-th/9906064].

\bibitem{msled}
C.P. Burgess, J. Matias and F. Quevedo, [hep-ph/0404135].
%

\bibitem{ABRS2}
A. Albrecht, C.P. Burgess, F. Ravndal and C. Skordis, {\it Phys.
Rev.} {\bf D65} (2002) 123506 (astro-ph/0107573).

\bibitem{AS}
A. Albrecht and C. Skordis, {\it Phys. Rev. Lett.} {\bf 84} 2076
(2000) (astro-ph/9908085).

\bibitem{LEDcosmobounds}
L.J. Hall and D. Smith, {\it Phys. Rev.} {\bf D60} (1999) 085008
(hep-ph/9904267);
%
M. Fairbairn, Phys.\ Lett.\ {\bf B508} (2001) 335-339
[hep-ph/0101131];
%
S. Hannestad, Phys.\ Rev.\ {\bf D64} (2001) 023515
[hep-ph/0102290].


\bibitem{LRTests}
C.M. Will, Living Rev.\ Rel.\ {\bf 4} (2001) 4, [gr-qc/0103036];
%
A. Aguirre, C.P. Burgess A. Friedland and D. Nolte, Class.\
Quant.\ Grav.\ 18 (2001) R223-- R232, [hep-ph/0105083];
%
G. Esposito-Farese, [gr-qc/0402007].

\bibitem{NLTests}
For a review with references, see E.G. Adelberger, B.R. Heckel and
A.E. Nelson, Ann.\ Rev.\ Nucl.\ Part.\ Sci.\ {\bf 53} (2003)
77--121, [hep-ph/0307284].

\bibitem{susyaddbounds2}
G. Azuelos, P.-H. Beauchemin and C.P. Burgess, ``Phenomenological
Constraints on Extra-Dimensional Scalars'', ATLAS publication
SN-ATLAS-2004-037, (hep-ph/0401125).
%
P.-H. Beauchemin, C.P. Burgess and G. Azuelos, ``Extra-Dimensional
Scalar Couplings to the Higgs Boson at the LHC'', preprint
McGill-04/12, ATLAS publication under review.


\end{thebibliography}
\end{document}